\documentstyle[aps,floats,prl,epsf]{revtex}

\def \s {\sigma}
\def \om {\omega}
\def \be{ \begin{eqnarray} }
\def \ee {\end{eqnarray}}
\def \bl {\begin{array}{ll}}
\def \el {\end{array}}
\def \ts {\textstyle}
\def \w {\uparrow}
\def \n {\downarrow}
\def \fb {{\bf f}}

\def \kb {{\bf k}}
\def \pb {{\bf p}}
\def \gb {{\bf g}}

\def \D {\Delta}
\def \bd {{\bf \delta}}
\def \sl {\sum \limits}
\def \l {\left}
\def \r {\right}
\def \N {{\ts \frac 1 N}}
\def \12 {{\frac 1 2}}

\newcommand{ \ve }{\varepsilon}
\newcommand{ \il }{\int \limits}
\newcommand{ \oil }{\oint \limits}
\def\bra#1{{\l< #1 \r|}}
\def\ket#1{{\l| #1 \r>}}
\def \ave#1 {{\l< #1 \r>}}
\def\abs#1{{\l| #1 \r|}}

\newcommand{\SR} {Sr_2RuO_4}
\def  \vol#1 { {\bf #1 } ~}
\def  \prb#1 {Phys. Rev. B { \bf #1 } ~}
\def  \prl#1 {Phys. Rev. Lett. { \bf #1 } ~}
\def  \jpc#1 {J. Phys. Cond. Matter { \bf #1 } ~}
\def  \nat#1 {Nature { \bf #1 } ~}
\begin{document}
\twocolumn[\hsize\textwidth\columnwidth\hsize\csname
@twocolumnfalse\endcsname
\title{Comparison of superconductivity in $Sr_2RuO_4$ and copper oxides}
\author{E.V.Kuz'min* and S.G.Ovchinnikov**}
\address{* Krasnoyarsk State University, Svobodnyi av. 74,
Krasnoyarsk 660062, Russia}
\address{** L.V.Kirensky Institute of Physics, Krasnoyarsk, 660036,
Russia}
\date{\today}
\maketitle
\begin{abstract}
To compare the superconductivity in strongly correlated electron
systems with the antiferromagnetic fluctuations in the copper
oxides and with the ferromagnetic fluctuations in $Sr_2RuO_4$ a
$t-J-I$ model is proposed. The antiferromagnetic coupling $J$
results in the superconducting state of $d_{x^2-y^2}$ symmetry and
the ferromagnetic coupling $I$ results in the spin-triplet
$p$-type state. The difference in the gap anisotropies provides
the large difference in $T_c$ values, for the typical values of
the coupling constants: $T_c \sim 1K$ for the ruthenate and $T_c
\sim 100K$ for the cuprates.\\
\vskip1mm {PACS numbers: 71.27.+a, 74.25}
\end{abstract}\vskip2pc]

More then a decade of intensive research of the cuprate
superconductors and related systems has raised fundamental
challenges to our understanding of the mechanism of
high-temperature superconductivity (SC). One of the most important
question is what is so specific in copper oxides, is it the unique
chemistry of the planar $Cu-O$ bond that determines the high value
of $T_c$ ? The discovery of SC in $Sr_2RuO_4$ with $T_c \sim 1K$
[1] is of a particular interest because it has a similar crystal
structure to the parent compound $La_2CuO_4$, of one of the best
studied families of the cuprate superconductors,
$La_{2-x}Sr_xCuO_4$, but has four valence electrons (for
$Ru^{4+}$) instead of one hole per formula unit. It is generally
believed that comparison of normal and SC properties of the
cuprates and the ruthenate will give more deeper understanding of
the nature of high-$T_c$ SC. While the normal state of doped
cuprates looks like almost antiferromagnetic Fermi-liquid [2], the
normal state of $Sr_2RuO_4$ is characterised by the strong
ferromagnetic fluctuations [3]. Properties of SC state are also
different: the singlet pairing with major contribution of the
$d_{x^2-y^2}$ symmetry was suggested for the cuprates [4], while
the triplet pairing with $p$-type symmetry similar to the $^3He~~
A_1$ phase is proposed for $Sr_2RuO_4$ [5]. The triplet SC in
$Sr_2RuO_4$ is induced by the ferromagnetic spin fluctuations [6].

 To compare the  SC in
$Sr_2RuO_4$ and cuprates we have proposed here a $t-J-I$ model
containing both an indirect antiferromagnetic coupling $J$ and a
direct ferromagnetic coupling $I$ between neighboring cations.
This model is based on the electronic structure calculations. An

important difference from the cuprates is that relevant orbitals
to the states near the Fermi energy are $Ru~~d \epsilon (d_{xy},
d_{yz}, d_{xz})$ and $O~~p \pi$, instead of $Cu~~d_{x^2-y^2}$ and
$O~~ p \sigma$ states. Due to $\s$-bonding in the cuprates a
strong $p-d$ hybridization takes place resulting in the strong
antiferromagnetic coupling $J$, a direct $d_{x^2-y^2}~~Cu-Cu$
overlapping is negligible. In $Sr_2RuO_4$ with $\pi$-bonding the
$Ru-O-Ru$ ~180-degree antiferromagnetic superexchange coupling is
weak [7] while a direct $d_{xy}~~Ru-Ru$ overlapping is not small.
That is why we add the Heisenberg type direct $Ru-Ru$ exchange
interaction to the Hamiltonian of the $t-J$ model. The strong
electron correlations are common features of the charge carriers
both in cuprates and $Sr_2RuO_4$ in our model. These correlations
for the cuprates are well known [8]. The importance of electron
correlations for $Sr_2RuO_4$ follows from the high value of the
effective mass of electrons in the $\gamma$-band obtained by the
quantum oscillations measurements [9].

We have found the different solutions for SC state in $t-J-I$
model: with the singlet $d$-type pairing governed by the
antiferromagnetic coupling $J$ and with the triplet $p$-type
pairing induced by the ferromagnetic coupling $I$. The equations
for $T_c$ in both states are similar. Nevertheless the same
absolute value of the coupling constants results in quite
different $T_c$ values, $T_c^{(p)} \sim 1K$ and $T_c^{(d)} \sim
100K$ for typical values of parameters. The gap anisotropy is
responsible for the large difference in the $T_c$ values. For the
$p$-type pairing the $\kb$-dependence of the gap provides
cancellation of the singular van-Hove contribution of the
two-dimensional density of states, while the $\kb$-dependence of
the $d$-type gap results in the significant contribution of the
van-Hove singularity.

 The Hamiltonian of the $t-J-I$ model is
written in the form \be \nonumber
 H= \sl_{\fb \s} (\ve-\mu) X_\fb^{\s \s} -t \sl_{\fb
{\bf \delta} \s} X_\fb^{\s 0} X_{\fb + \bd}^{0\s} + J \, \sl_{\fb
\bd} K^{(-)}_{\fb,\fb+\bd}- \ee \be\label{ham} I \, \sl_{\fb \bd}
K^{(+)}_{\fb,\fb+\bd}~, \ee \be \nonumber
 K^{(\pm )}_{\fb \gb}={\vec S_\fb} \cdot {\vec
S_\gb} \pm \frac 1 4
 n_\fb  n_\gb~,~~~ X_{\fb}^{\w \w} +X_{\fb}^{\n \n}+X_{\fb}^{00} =1.
\ee Here the Hubbard X-operators $X_\fb^{pq}=\ket{p} \bra{q}$ are
determined in the reduced Hilbert space containing empty states
$\ket{0}$ and single-occupied states $\ket{\s}$, $\s =\w$ or $\s
=\n$. The X-operators algebra exactly takes into accounts the
constrain condition that is one of the important effects of the
strong electron correlations. The $\vec S_\fb$ and $n_\fb$
operators in (\ref{ham}) are the usual spin and number of
particles operators at the site $\fb$, $\bd$ is the vector between
n.n.. For the cuprates $J \gg I$, but for $Sr_2RuO_4$ $I \gg J $.

To get SC the copper oxides should be doped while $ \SR $ is
self-doped. According to the band structure calculations [10] the
electron $ \alpha  $-band in $\SR $ is half-filled, the hole $
\beta $-band has $n_0=0.28$ holes and the electron $ \gamma $-band
with $d_{xy}$ contribution is more then half-filled, $n_ \gamma
=1+n_0$. The strong electron correlations split the $ \gamma
$-band into filled lower Hubbard band (LHB) with $n_e=1$and
partially filled upper Hubbard band (UHB) with the electron
concentration $n_e =n_0$. We use the hole representation where the
electron UHB transforms in the hole LHB with hole concentration
$n_h =1-n_0$. All other bands ( $\alpha$ and $\beta$ ) are treated
here as an electron reservoir. Observation of a square flux-line
lattice in $\SR$ allows to suggest that SC resides mainly on the $
\gamma $ band [11]. For the cuprates the quasiparticle is a hole
in the electron LHB with the electron concentration $n_e =1-n_0$,
for $La_{2-x}Sr_xCuO_4$ $n_0=x$.
\begin{figure}[t]
\epsfxsize=8cm \epsffile{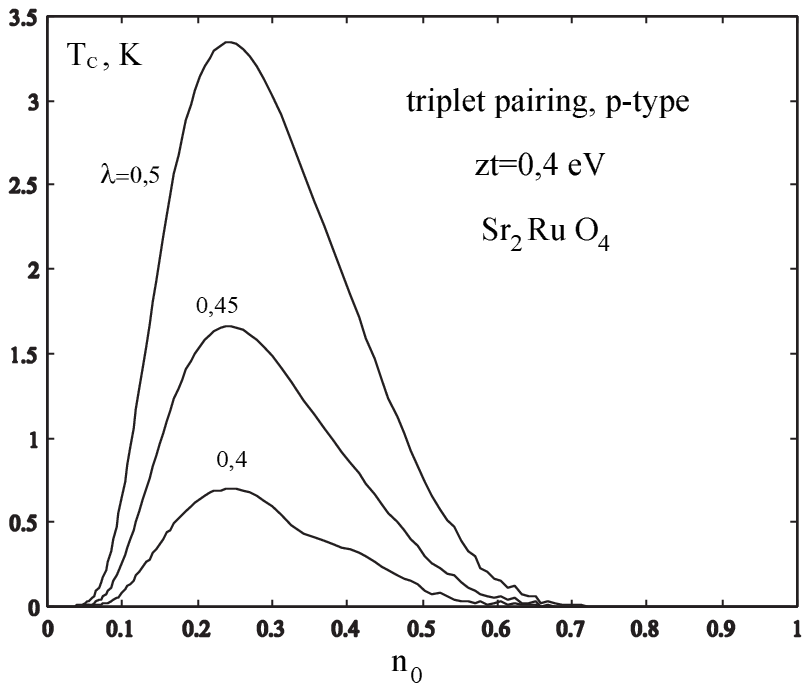} \caption{
 Concentration dependence of $T_c$ for the triplet pairing
of the $p$-type. } \label{fig1}
\end{figure}
\begin{figure}[h]
\epsfxsize=8cm \epsffile{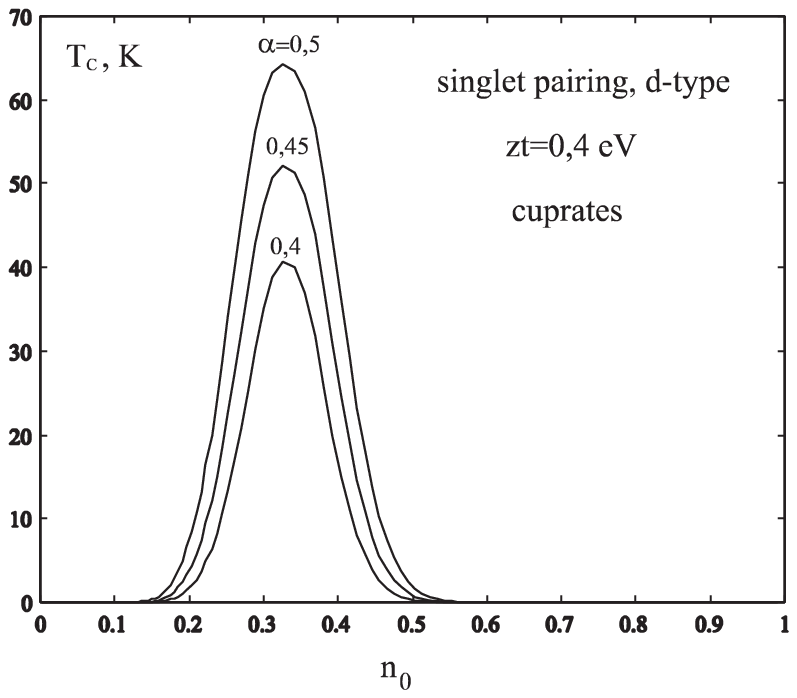} \caption{
 Concentration dependence of $T_c$ for the singlet pairing
of the $d$-type. }\label{fig2}
\end{figure}
There are many ways to get the mean-field solutions for SC state,
we have used the irreducible Green function method [12] projecting
the higher-order Green functions onto subspace of normal $\l< \l<
X_{\kb}^{0 \s} \mid  X_{\kb}^{\s 0} \r> \r>$ and abnormal $\l< \l<
X_{-\kb}^{-\s 0} X_ {\kb}^{\s 0} \r> \r>$ Green functions coupled
via the Gorkov system of equations. Three different solutions have
been studied: singlet $s$- and $d$- and triplet $p$-types. The gap
equation has the form for $s$-state \be \label{eqS} 1 = \N
\sl_{\pb} \frac {2\om_\pb +(2g-\lambda)\om^2_\pb}{2E_{\pb0}}
\,\tanh \l( \frac {E_{\pb0}}{2\tau} \r) \ee and for the
$p,d$-states
 \be \label{eqL} \frac
{1}{ \alpha _l} =
  \frac 1 N \sl_\pb \frac {\psi^{2}_{l}(\pb)}{2E_{\pb l}}
\, \tanh \l( \frac {E_{\pb l}}{2 \tau} \r )~.  \ee  Here for $s,
p,d$-states  \be \nonumber E_{\pb l}= \sqrt{c^2(n_0)(\om_\pb-m)^2
+ \abs{\D_{\pb l}}^2}~,\ee \be \nonumber m=\l. \l[\frac {\mu-
\epsilon }{zt} +(g+\lambda)\frac {1-n_0}{2} \r] \r/ c(n_0)~, \ee
where $\om_\pb = - \gamma _\pb = -(1/z) \sl_{\bd}\exp(i \pb \bd)$,
 $m$ is a
dimensionless chemical potential, $c(n_0)=(1+n_0)/2$, $g=J/t,
\lambda=I/t$ and $\tau=k_B T/zt$ is a dimensionless temperature.
The SC gaps are equal to \be\nonumber \D_{\kb 0}= [2+(2g-\lambda)
\om_\kb]~\D_0~, \D_0 = \N \sl_{\pb} \om_{\pb} B_{\pb}/c(n_0)~,\ee
\be \label{gapS} B_\pb = \l< X_{-\pb}^{0\n} X_{\pb}^{0\w} \r>~ \ee
for $s$-type $(l=0)$ and
 \be \label{gapL} \D_{\kb l} = \alpha_l
\psi_l(\kb)
 \frac 1 N \sl_\pb \psi_l (\pb)B_{\pb}~/c(n_0) \ee for the $p$- and $d$-
 states. The coupling constants and the gap anisotropy in the $l$-th
channel are given by  \be \label{P}
 \alpha _p=\lambda~,~~\psi_{p}(\kb)=\12 (\sin\,k_x + \sin\,k_y)~,\ee
\be \label{D}  \alpha _d=(2g-\lambda)~,~~\psi_{d}(\kb)= \12
(\cos\,k_x -\cos\,k_y)~. \ee Here we have considered only the
two-dimensional square lattice with the lattice parameter $a=1$.
The equation for the chemical potential has the form \be
\label{conc} 1-n_0= \N \sl_{\kb \s} \l< X_{\kb}^{\s 0} X_{\kb}^{0
\s} \r>~. \ee  An important effect of the strong electron
correlation is the constrain condition excluding double-occupied
states  \be \label{sumr} \N \sl_\kb B_\kb~ = \N \sl_\kb \l<
X_{-\kb}^{0 \n} X_{\kb}^{0 \w}\r>=0~. \ee

The first term in (\ref{eqS}) for the singlet $s$-type pairing
parameter  is proportional to $2tz$ and appears due to the
kinematic mechanism of pairing [13]. The $s$-type solution does
not satisfy to the constrain condition (\ref{sumr}) [14] while for
$p$- and $d$-type it is fulfilled. The equation for $T_c$ in $p$-
and $d$-states is given by \be \label{TC} \frac{2c(n_0)}{ \alpha
_l} = \N \sl_\pb \frac {\psi_l^{2}(\pb)}{\abs{\om_{\pb} -
m}}~\tanh \l( \frac {c(n_0) \abs{\om_{\pb} - m}}{2\tau_c^{(l)}}
\r)~. \ee The same equation for the $d_{x^2-y^2}$-pairing has been
derived by the diagram technique for the $t-J$ model [15].

At the numerical solution of the equations (\ref{TC}) more then
$10^6$ points of the Brillouine zone have been taken. Results of
$T_c(n_0)$ computations are shown in the Fig.1 and 2 for several
values of the coupling constants $ \alpha _l$. These results have
revealed the remarkable difference in $T_c$ values: $T_c^{(p)} \ll
T_c^{(d)}$ when $ \alpha _p =  \alpha _d$. The moderate values of
$ \alpha  \approx 0.4-0.5$ and $zt \approx 0.5eV$ result in
$T_c^{(p)} \sim 1K, T_c^{(d)} \sim100K$.

It is clear from equations (\ref{P}),(\ref{D}) that the $p$-type
SC is formed by the ferromagnetic interaction, that is the case of
$\SR$, and the $d$-type SC is induced by the antiferromagnetic
interaction in copper oxides. To understand why $T_c^{(d)} \gg
T_c^{(p)}$ we have analysed the eqn. (\ref{TC}) analytically.
Using integration over the constant energy surfaces $\om_\kb =
\om$ it can be rewritten like  \be \label{Int} \frac {2c(n_0)}{
\alpha _l}=\il_{-1}^{+1}\,
 \frac {\psi^2_{l}(\om)}{\abs{\om -m}}\,
\tanh \l(\frac {c(n) \abs{\om -m}}{2\tau_c} \r)\, d\om~, \ee \be
\label{PSI} \psi^2_{l}(\om)=\frac{1}{(2 \, \pi)^2} \, \oil_{\l(
\s_\om \r)} \, \frac {\psi^2_{l}(\kb)}{ \abs{\nabla_\kb \om_\kb}}
\, d \s_\om . \ee The sum rule for the $\psi^2_{l}(\om)$ functions
is the same for $l=p$ and $l=d$ : \be \label{NORM} \N \sl_\kb
\psi^2_{l}(\kb) = \il_{-1}^{+1} \psi^2_{l}(\om)\,d\om\,= 1/4 . \ee
For the $p$-state \be   \label{Pt}
\frac{\psi^2_{p}(\kb)}{\abs{\nabla_\kb \om_\kb}}= \abs{\nabla_\kb
\om_\kb} + \frac{\sin \, k_x \, \, \sin \, k_y}{\sqrt{\sin^2\,
k_x+\sin^2 \, k_y}} \ee the second term in (\ref{Pt}) gives zero
contribution to the integral (\ref{PSI}) and $\psi^2_{p}(\om)$ is
rather small with smooth energy dependence, $\psi^2_{p}(\om)
\approx (2/\pi^2)(1-\abs{\om}^{1.61})$. For the $d$-state \be
\label{Dt} \frac {\psi^2_{d}(\kb)}{\abs{\nabla_\kb \om_\kb}}=
\frac 12 \frac{(\cos \, k_x -  \cos \, k_y)^2}{\sqrt{\sin^2\,
k_x+\sin^2 \, k_y}} \ee has the same singularity as the van-Hove
singularity in the density of states $\rho(\om)$. The result of
calculation is \be \nonumber \psi^2_{d}(\om) =
(1-\om^2)\rho(\om)-2\psi^2_{p}(\om)\, ,\ee \be
\label{rho}\rho(\om) = \frac {1}{\pi} -\l(\frac 1 2 -\frac
{1}{\pi}\r) \ln(\abs{\om}). \ee The comparison of
$\psi^2_{p}(\om)$ and $\psi^2_{d}(\om)$ has shown that the
van-Hove singularity is cancelled in the $p$-state and does not
cancelled in the $d$-state (Fig 3).
\begin{figure}[h]
 \epsfxsize=8cm \epsffile{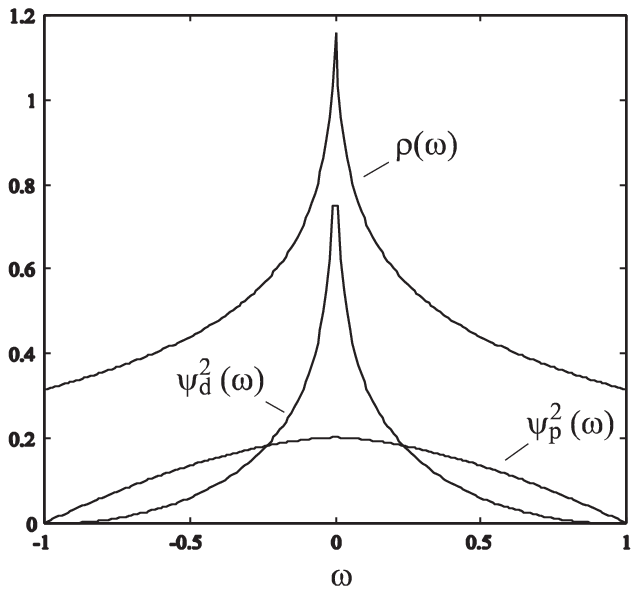} \caption{ Energy
dependence of the effective gap anisotropy $\psi^2_p$ and
$\psi^2_d$ and single-particle density of states $\rho(\omega)$ in
square lattice. }\label{fig3}
\end{figure}

In conclusion we have presented the model of strongly correlated
electrons in two dimensional lattice that allows to consider the
cuprates ($J \gg I$) and $\SR $ $(J \ll I)$ on the same footing.
The singlet SC  in the $s$-state is absent in the strong
correlation limit, the triplet $p$-pairing occurs due to the
ferromagnetic fluctuations and the singlet $d$-pairing is induced
by the antiferromagnetic fluctuations. The reason why $T_c$ in the
cuprates is much higher then in $\SR$ is the different gap
anisotropies. For the $p$-state $\kb$-dependence of the gap
results in the cancellation of the van-Hove singularity  while for
the $d$-state the gap anisotropy permits large van-Hove
singularity contribution  in the equation for $T_c$. For the
question what is so specific in the copper oxides for high-$T_c$
superconductivity the possible answer may be as follows: it is the
planar $Cu-O~~\s$-bonding resulting in the strong
antiferromagnetic $Cu-Cu$ interaction, that induced the singlet
pairing with $d_{x^2-y^2}$ symmetry.

We thank N.M.Plakida for useful discussions and I.O.Baklanov for
numerical calculations. The work was supported by the Russian
Federal Program "Integration of high school education and
science", grant N69.

\begin{center}
*****
\end{center}

*Present address: L.V.Kirensky Institute of Physics, Krasnoyarsk,
660036, Russia.

Electronic address: sgo@post.krascience.rssi.ru

\begin{center}
- - - - -
\end{center}

 [1] Y.Maeno, H.Hasimoto, K.Yoshida, S.Nisshizaki, T.Fujita,
F.Lichtenberg, \nat{372}, 532 (1994).

[2] D.Pines, Physica B \vol {163}, 78 (1990).

[3] T.Oguchi, \prb{51}, 1385 (1995).

[4] N.E.Bickers, D.J.Scalapino, R.T.Scaletar,\\ Int. J. Mod. Phys.
\vol {B1}, 687 (1987).

[5] T.M.Rice, H.Sigrist, \jpc{7},\\ L 643 (1995).

[6] I.I.Mazin, D.J.Singh, \prl{79}, \\ 733 (1997).

[7] J.B.Goodenough, {\it Magnetism and chemical bond} (John Wiley
and Sons, N.Y.-London, 1963).

[8] E.Dagotto, Rev. Mod. Phys. \vol{66}, 763 (1994).

[9] A.P.Mackenzie, S.R.Julian, A.J.Diver {\it et al},\\ \prl{76},
3786 (1996).

[10] D.J.Singh, \prb{52}, 1358 (1995).

[11] T.M.Riseman, P.G.Kealey, E.M.Forgan,\\ A.P.Mackenzie,
L.M.Galvin, A.W.Tyger, S.L.Lee,\\ C.Ages, D.McK.Paul,
C.M.Aegerter, R.Cubitt, Z.Q.Hao, T.Akima, Y.Maeno, Nature (London)
\vol{396}, 242 (1998).

[12] S.V.Tyablikov, {\it Methods of quantum theory of magnetism}
(2-nd edition, Moscow, Nauka, 1975).

[13] R.O.Zaitsev, V.A.Ivanov, Fiz. tverdogo tela \\(Sov. Solid
State Physics) \vol{29}, 2554 (1987).

[14] N.M.Plakida, V.Yu.Yushanhai, I.V.Stasyuk,\\ Physica C\vol{
160}, 80 (1989).

[15] Yu.A.Izyumov, M.I.Katsnelson, Yu.N.Skryabin, {\it Magnetism
of itinerant electrons} (Moscow, Nauka, 1994).

~~~~~~~~~~~~~~~~~~~~~~~~~~~~~~~~~~~~~~~~~~~~~~~~~~~~~~~~~~~~~~~~~~~~~~~
~
\begin{center}
*****
\end{center}

~~~~~~~~~~~~~~~~~~~~~~~~~~~~~~~~~~~~~~~~~~~~~~~~~~~~~~~~~~~~~~~~~~~~~~~
\clearpage

\end{document}